\documentstyle[aps,prc]{revtex}
\begin{document}
\draft
\title{Finite range effects of nuclear force in intermediate energy heavy ion
collisions}
\author{Lie-Wen Chen$^{1,2,3}$, Feng-Shou Zhang,$^{1,2,4}$ Wen-Fei Li$^{1,2}$}
\address{$^1$ Center of Theoretical Nuclear Physics, National Laboratory of Heavy Ion%
\\
Accelerator, Lanzhou 730000, China\\
$^2$ Institute of Modern Physics, Academia Sinica, P.O. Box 31, Lanzhou
730000, China\\
$^3$ Department of Applied Physics, Shanghai Jiao Tong University, Shanghai
200030, China\\
$^4$CCAST (World Laboratory), P.O. Box 8730, Beijing 100080, China}
\maketitle

\begin{abstract}
Within the framework of an isospin-dependent quantum molecular dynamics
model, the zero-range 2-body part of the Skyrme interaction is replaced by a
finite-range Gaussian 2-body interaction. From the transverse momentum
analysis in the reaction of system $^{93}$Nb + $^{93}$Nb at energy of 400
MeV/nucleon and impact parameter $b$=3 fm, it is shown that the finite-range
nuclear force enhances the transverse momentum of the reaction system and it
can partly replace the momentum dependent part of the nucleon-nucleon
interaction.
\end{abstract}


\section{Introduction}

The heavy-ion transport theory is one of important research frontiers in the
theoretical nuclear physics in the last decade. Since 1980s, the beam energy
has gone up to intermediate and high energy regions and the high temperature
and high dense nuclear matter could be formed in the heavy-ion collisions
(HIC's), through which one could investigate the nuclear matter properties
at extreme conditions. Correspondingly, theoretical nuclear physicists have
established some microscopic heavy-ion transport theory, among which the
Bolztmann-Uehling-Uhlenbeck (BUU) equation\cite{be88}, Boltzmann Langevin
(BL) equation\cite{ab96}, Quantum Molecular Dynamics (QMD) model\cite{ai91},
Fermionic Molecular Dynamics (FMD) model\cite{fe90}, and Antisymmetric
Molecular Dynamics (AMD) model\cite{on92}, have been used extensively and
successfully. Though these models have respective advantages at different
aspects, they have a common character, namely, the zero-range Skyrme-like
effective interaction being used in all these models. It is well known,
however, that the nuclear force is attractive at long range and repulsive at
short range. Hence, the finite-range forces may be as more ``realistic''
than zero-range Skyrme-like forces. With the zero-range forces, some
momentum-dependent terms are quadratic or constant leading to irrealistic
behavior while those terms vanish with finite-range forces even though
relative momenta go high\cite{id94}. At usual relative momenta, below 2$k_F$%
, the behavior of zero-range forces are still acceptable but it is always
interesting to study the features of HIC's with finite-range nuclear forces%
\cite{se89}.

Based on an isospin-dependent QMD model\cite{ch97,ch98,ch99,zh99,ch00} which
includes the symmetry potential, isospin dependent nucleon-nucleon ({\sl N-N}%
) collisions, Coulomb potential for protons, and isospin dependent Pauli
blocking effects, the zero-range 2-body Skyrme interaction is replaced by a
finite-range Gaussian interaction. As the first investigation of the
finite-range effects, we study the collective flow in reaction $^{93}$Nb + $%
^{93}$Nb at energy of 400 MeV/nucleon and impact parameter $b$=3 fm since
the finite range of nuclear force represents momentum-dependence of nuclear
force in the momentum space\cite{ri80} and the collective flow is sensitive
to the momentum dependent interaction. The calculated results indicate that
the finite-range 2-body interaction actually enhances the transverse
momentum, which implies that finite-range nuclear force could replace partly
the momentum-dependent part of the {\sl N-N} interaction and therefore by
using the finite-range {\sl N-N} interaction the momentum dependence of the
nuclear forces is considered naturally.

\section{Model and method}

In the QMD model, nucleon $i$ is represented by a Gaussian form of wave
function: 
\begin{equation}
\Psi _i({\bf r},t)=\frac 1{(2\pi L)^{3/4}}e^{-[{\bf r}-{\bf r}%
_i(t)]^2/(4L)}e^{i{\bf p}_i\cdot {\bf r}/\hbar }\text{.}
\end{equation}
Performing Wigner transformation for Eq. (1), one can get the nucleon's
Wigner density distribution in phase space:

\begin{equation}
f_i({\bf r},{\bf p},t)=\frac 1{\pi \hbar ^3}exp[-\frac{({\bf r}-{\bf r}%
_i(t))^2}{2L}-\frac{({\bf p}-{\bf p}_i(t))^2\cdot 2L}{\hbar ^2}]\text{,}
\end{equation}
where the ${\bf r}_i$ and ${\bf p}_i$ represent the mean position and
momentum of the $i$th nucleon, respectively, the $L$ is the so-called
Gaussian wave packet width (here $L$=2.0 fm$^2$). In the QMD model, then the
total interaction potential can read

\begin{equation}
U^{tot}=U^{(2)}+U^{(3)},
\end{equation}
and the two-body and three-body potentials, $U^{(2)}$ and $U^{(2)}$ can be
given, respectively, by

\begin{equation}
U^{(2)}=\sum\limits_{i(\neq j)}U_{ij}^{(2)}=\sum\limits_{i(\neq j)}\int f_i(%
{\bf r},{\bf p},t)V_{ij}({\bf r},{\bf r}^{^{\prime }})f_j({\bf r}^{\prime },%
{\bf p}^{\prime },t)d{\bf r}d{\bf p}d{\bf r}^{\prime }d{\bf p}^{\prime },
\end{equation}
and 
\begin{eqnarray}
U^{(3)} &=&\sum\limits_{i\neq j,i\neq k,j\neq k}U_{ijk}^{(3)}  \nonumber \\
&=&\sum\limits_{i\neq j,i\neq k,j\neq k}\int f_i({\bf r},{\bf p},t)f_j({\bf r%
}^{\prime },{\bf p}^{\prime },t)f_k({\bf r}^{\prime \prime },{\bf p}^{\prime
\prime },t)V_{ijk}({\bf r},{\bf r}^{\prime },{\bf r}^{\prime \prime })d{\bf r%
}d{\bf p}d{\bf r}^{\prime }d{\bf p}^{\prime }d{\bf r}^{\prime \prime }d{\bf p%
}^{\prime \prime }.
\end{eqnarray}

The two-body interaction $V_{ij}$ includes the local two-body Skyrme
interaction $V_{ij}^{loc}$, the Yukawa (surface) interaction $V_{ij}^{Yuk}$,
the symmetry energy part $V_{ij}^{sym}$, and the Coulomb interaction $%
V_{ij}^{Coul}$. Correspondingly, the total two-body potential in the QMD
model is as follows,

\begin{equation}
U^{(2)}=U_{(2)}^{loc}+U^{Yuk}+U^{sym}+U^{Coul}=\sum\limits_{i(\neq
j)}(U_{ij}^{loc}+U_{ij}^{Yuk}+U_{ij}^{sym}+U_{ij}^{Coul}),
\end{equation}
for the forms of $U^{Yuk}$, $U^{sym}$, and $U^{Coul}$, one is refereed to
Refs.\cite{ch97,ch98,ha89,Bass95,ha98}. Refs.\cite{ha89,Bass95,ha98} give a
detailed description for the Isospin-QMD (IQMD) model. In the present
isospin-dependent QMD model, however, Pauli blocking of neutron and proton
is treated respectively, namely, the Pauli blocking is isospin dependent.
The two-body Skyrme interaction is given as following form 
\begin{equation}
V_{ij}^{loc}=t_1\delta ({\bf r}_i-{\bf r}_j).
\end{equation}
From Eqs. (4) and (7), one can get the two-body Skyrme potential

\begin{equation}
U_{(2)}^{loc}=\sum\limits_{i(\neq j)}U_{ij}^{(loc)}=\sum\limits_{i(\neq
j)}t_1\frac 1{4\pi L^{3/2}}exp[-\frac{({\bf r}_i-{\bf r}_j)^2}{4L}].
\end{equation}
Similarly, from the three-body Skyrme interaction

\begin{equation}
V_{ijk}^{(3)}=t_2\delta ({\bf r}_i-{\bf r}_j)\delta ({\bf r}_i-{\bf r}_k),
\end{equation}
one can get the three-body Skyrme potential

\begin{eqnarray}
U_{(3)}^{loc} &=&\sum\limits_{i\neq j,i\neq k,j\neq k}U_{ijk}^{(loc)} 
\nonumber \\
&=&\sum\limits_{i\neq j,i\neq k,j\neq k}t_2\frac 1{3^{3/2}(2\pi L)^3}exp[-%
\frac{({\bf r}_i-{\bf r}_j)^2+({\bf r}_i-{\bf r}_k)^2+({\bf r}_j-{\bf r}_k)^2%
}{6L}].
\end{eqnarray}

If one defines the interaction density,

\begin{equation}
\rho _{int}^i(r_i)=\frac 1{4\pi L^{3/2}}\sum\limits_{j\neq i}exp[-\frac{%
(r_i-r_j)^2}{4L}],
\end{equation}
then Eqs. (8) and (10) can be approximately combined into

\begin{equation}
U^{loc}=\alpha (\frac{\rho _{int}}{\rho _0})+\beta (\frac{\rho _{int}}{\rho
_0})^\gamma ,
\end{equation}
where the $\rho _0$ is the normal nuclear matter density, 0.16 fm$^{-3}$.

For the momentum dependent interaction $V_{ij}^{MDI}$, we make use of the
real part of the optical potential parametrized in Ref. \cite{ai87} as
follows

\begin{equation}
V_{ij}^{MDI}=\delta ln^2[\varepsilon ({\bf p}_i-{\bf p}_i)^2+1]\delta ({\bf r%
}_i-{\bf r}_j).
\end{equation}
The parameters of Eqs. (12) and (13) are given in Table I, from which one
can see two kinds of equations of state (EOS) are commonly used. One is the
so-called hard EOS (H, HM) with an incompressibility of $K$=380 MeV, and the
other is the soft EOS (S, SM) with an incompressibility of $K$=200 MeV\cite
{ai91}. The $M$ refers to the inclusion of the momentum dependent
interaction.

If the 2-body Skyrme interaction, Eq. (7), is replaced by a finite-range
Gaussian, i.e.,

\begin{equation}
V_{ij}^{loc}=t_1\frac 1{\Delta ^3\pi ^{3/2}}exp[-\frac{({\bf r}_i-{\bf r}%
_j)^2}{\Delta ^2}],
\end{equation}
where the $\Delta $ represents the finite-range parameter of {\sl N-N}
interaction, then one can get the Gaussian 2-body effective potential in the
QMD model

\begin{equation}
U_{(2)}^{Gau}=\sum\limits_{i(\neq j)}U_{ij}^{(Gau)}=\sum\limits_{i(\neq
j)}t_1\frac 1{[4\pi (L+\frac{\Delta ^2}4)]^{3/2}}exp[-\frac{({\bf r}_i-{\bf r%
}_j)^2}{4(L+\frac{\Delta ^2}4)}].
\end{equation}

By comparing Eq. (15) with Eq. (8), one can find that the Gaussian 2-body
effective potential can be reached only through replacing the $L$ by $L+%
\frac{\Delta ^2}4$ in the 2-body Skyrme potential, namely, Eq. (8). For the
zero-range Skyrme two-body interaction, Eq. (7), the short range repulsive
core is not considered. Physically, the Eq. (14) is equivalent to simulate
the short range repulsive core and therefore $\Delta $ should have the same
order as radius of the repulsive core, namely, $\Delta \approx 0.4\sim 0.6$
fm. The long range attractive part of the 2-body {\sl N-N} interaction
mainly comes from the Yukawa interaction. In order to observe the effects of 
$\Delta $ on nuclear force, we display in Fig. 1 the finite-range Gaussian
2-body force as a function of the distance between the two nucleons for $%
\Delta =0.0$, $0.3$, $0.5$ and $0.7$ fm. Here we set $t_1=356$ MeV$\cdot $fm$%
^3$ and it should be noticed that the case of $\Delta =0.0$ corresponds
exactly to the case of zero-range Skyrme two-body force since Eq. (7) is the
limitation of Eq. (14) as $\Delta \rightarrow 0$. It is indicated in Fig. 1
that the 2-body nuclear force is attractive and becomes weak gradually with
increment of $\Delta $, which imply the finite-range enhances the repulsive
effect of nuclear force. This phenomenon is easy to understand since the
finite range of nuclear force means momentum-dependence of nuclear force in
the momentum space.

Similarly, the zero-range 3-body Skyrme interaction, Eq. (9), can be
replaced by finite-range Gaussian 3-body interaction, i.e.,

\begin{equation}
V_{ijk}^{(3)}=t_2(\frac 1{\Delta ^2\pi })^3exp[-\frac{({\bf r}_i-{\bf r}%
_j)^2+({\bf r}_i-{\bf r}_k)^2}{\Delta ^2}],
\end{equation}
after tedious algebra, one can get the finite-range Gaussian 3-body
effective potential in the QMD model, i.e.,

\begin{eqnarray}
U_{(3)}^{Gau} &=&\sum\limits_{i\neq j,i\neq k,j\neq
k}U_{ijk}^{(Gau)}=\sum\limits_{i\neq j,i\neq k,j\not \neq k}\frac{t_2}{%
3^{3/2}[2\pi \sqrt{(L+\frac{\Delta ^2}6)(L+\frac{\Delta ^2}2)}]^3}  \nonumber
\\
&&\ \times exp[-\frac{(\Delta ^2+2L)[({\bf r}_i-{\bf r}_j)^2+({\bf r}_i-{\bf %
r}_k)^2]+2L({\bf r}_j-{\bf r}_k)^2}{12(L+\frac{\Delta ^2}6)(L+\frac{\Delta ^2%
}2)]}].
\end{eqnarray}
Obviously, this expression is very complicated and has not the simplicity of
Eq. (15). In fact, neglecting the $\Delta ^4$ and its higher order terms,
Eq. (17) can be also simplified as following form, 
\begin{eqnarray}
U_{(3)}^{Gau} &=&\sum\limits_{i\neq j,i\neq k,j\neq
k}U_{ijk}^{(Gau)}=\sum\limits_{i\neq j,i\neq k,j\not \neq k}\frac{t_2}{%
3^{3/2}[2\pi (L+\frac{\Delta ^2}3)]^3}  \nonumber \\
&&\ \times exp[-\frac{({\bf r}_i-{\bf r}_j)^2+({\bf r}_i-{\bf r}_k)^2+({\bf r%
}_j-{\bf r}_k)^2}{6(L+\frac{\Delta ^2}3)}].
\end{eqnarray}
Surprisingly, by comparing Eq. (18) with Eq. (10), one can also find that
the Gaussian 3-body effective potential can be reached only through
replacing simply the $L$ by $L+\frac{\Delta ^2}3$ in the 3-body Skyrme
potential, namely, Eq. (10). Generally, the zero-range 3-body part of Skyrme
interaction can be regarded as a good approximation for the 3-body part of 
{\sl N-N} effective interaction. Therefore, we only insert the finite-range
2-body part into the QMD code in the present work.

It should be noted that the replacing of the $L$\ by $L+\frac{\Delta ^2}4$\
or $L+\frac{\Delta ^2}3$\ is only for the two-body or three-body effective
potential in the QMD model and this replacing dose not affect the other
components of the QMD model. Therefore, it is not simply to modify the width
of the Gaussian wave packet in the QMD model. The change of the two-body or
three-body effective potential only comes from the using of the two-body or
three-body finite-range Gaussian interaction (Eq. (14) or (16) ) in stead of
the two-body or three-body zero-range Skyrme interaction (Eq. (7) or (9) ).
In fact, the present method is very similar with that of using the
well-known finite-range Gogny interaction where the finite range is given by
the superposition of two Gaussian function with different ranges and
spin-isospin mixtures.

\section{Results and discussions}

Within the framework of the above modified QMD model, the transverse
momentum is calculated for the reaction of $^{93}$Nb + $^{93}$Nb at energy
of 400 MeV/nucleon and impact parameter $b$=3 fm for five different
potential parameter sets, namely, hard EOS (H), soft EOS (S), soft EOS with
momentum dependent interaction (SM), soft EOS with $\Delta =0.1$ (S 0.1),
and soft EOS with $\Delta =0.5$ (S 0.5). Fig. 2 gives the time evolution of
transverse momentum per nucleon for the above five cases. From Fig. 2 one
can see clearly that at the initial stage of the collisions the transverse
momentum is negative for all cases except for the case with SM, which is
easy to understand since for the cases of H, S, S 0.1, and S 0.5, at the
initial stage of collisions the attractive nuclear mean field is dominant
and consequently the transverse momentum is negative. However, for the case
of SM, the initial large relative momentum leads to a large repulsive
momentum dependent potential which balances the attractive part of nuclear
mean field.

With time evolution, the projectile and target begin to collide, compress,
and expand, and the positive transverse momentum is generated. It is shown
in Fig. 2 that the transverse momenta have saturated after 40 fm/c for all
five cases. In the saturated region, one can see that the largest transverse
momentum is observed in the case of H and the smallest in the case of S. The
transverse momentum in the case of S 0.1 is little larger than that in the
case of S. Comparing to the case of S 0.1, the case of S 0.5 has a larger
transverse momentum which is still smaller than that in the case of SM by an
amount of 8 MeV/c. These features indicate that the potential parameter set
of soft EOS with $\Delta =0.5$ could replace partly that of soft EOS with
the momentum dependent interaction, which implies that the finite-range
Gaussian two-body nuclear force could replace partly the momentum dependent
part of the {\sl N-N} interaction.

In order to compare the present calculated results with the experimental
data, we display in Fig. 3 the rapidity distribution of the transverse
momentum for above five potential parameter sets. The big open circles
included in Fig. 3 represent the experimental data which are from the Ref.%
\cite{ri85}. The small open circles linked by lines represent the calculated
results. Fig. 3 (a) shows the calculated results for the case of H, which
indicates that the calculated results are in agreement with the experimental
data. Fig. 3 (b) corresponds to the case of SM and it is shown that the
calculated results are in better agreement with the experimental data. In
Fig. 3 (c) the calculated values linked by solid line and dashed line
correspond, respectively, to the cases of S and S 0.1, which are not in good
agreement with the experimental data. In the case of S 0.5, Fig. 3 (d)
indicates that the calculated results are basically in agreement with the
experimental data. These features imply that all the calculated results with
potential parameter sets H, SM, and S 0.5 are in agreement with the
experimental data while those with potential sets S and S 0.1 fail to
reproduce the experimental data. Therefore, the finite-range parameter $%
\Delta =0.5$\ fm is very reasonable to model the finite-range effect of
nuclear force.

Fig. 4 displays the time evolution of transverse momentum per nucleon for
free nucleons (A=1, solid circles in Fig. 4) and light fragments (A=2$\sim $%
4, open circles in Fig. 4) for four different cases, namely, Figs. 4 (a),
(b), (c), and (d) correspond, respectively, to the cases of H, S, SM, and S
0.5. In this paper, we construct clusters in terms of the so-called
coalescence model\cite{zh95} with coalescence parameters $R_0=3.5$ fm and $%
P_0=260$ MeV/c. One should note that the mass dependence of the transverse
momentum shown in Fig. 4 demonstrates the well known increase in magnitude
for heavier fragments\cite{bo91,hu91,pe90}. This phenomenon may be because
most nucleons are emitted by the hard stochastic collisions and hence the
effect of the mean field is largely erased in the nucleon flow. This
argument suggests that the flow of composite fragment carries more direct
information of the nuclear EOS than the nucleon flow. From Fig. 4 one can
also find that the finite-range Gaussian two-body nuclear force enhances the
transverse momenta of fragments and could replace partly the momentum
dependent part of the {\sl N-N} interaction.

\section{Conclusions}

A finite-range Gaussian two-body interaction is applied in stead of the
zero-range two-body part of the Skyrme interaction in an isospin-dependent
QMD model to simulate the transverse momentum in the reaction of system $%
^{93}$Nb + $^{93}$Nb at energy of 400 MeV/nucleon and impact parameter $b$=3
fm. The calculated results show that the finite-range nuclear force enhances
the transverse momentum of the reaction system and it can partly replace the
momentum dependent part of the {\sl N-N} interaction. The calculated results
with potential parameter set of soft EOS with finite-range parameter $\Delta
=0.5$ are basically in agreement with the experimental data. Meanwhile, it
is shown that the transverse momentum of light fragments is greater than
that of free nucleon, which agrees with the results of many experiments.

This work is a primary investigation of the finite-range effects of nuclear
force and further consideration is worth performing. For example, the
two-body forces adopt the finite-range Gogny forces and meanwhile the
finite-range three-body force should be included. In addition, it is very
interesting to explore the finite-range effects of other physical phenomena.

\section{Acknowledgments}

This work was supported by the National Natural Science Foundation of China
under Grant NOs. 19875068 and 19847002, the Major State Basic Research
Development Program under Contract NO. G2000077407, and the Foundation of
the Chinese Academy of Sciences.

\newpage

\section*{Table Captions}

\begin{description}
\item[Table I]  The parameter sets of Eqs. (2) and (3). The S and H refer to
the soft and hard equations of state, the $M$ refers to the inclusion of
momentum dependent interaction, and the $K$ refers to the incompressibility.
\end{description}

\begin{center}
{\small 
\begin{tabular}{ccccccc}
\hline\hline
& K(MeV) & $\alpha $(MeV) & $\beta $(MeV) & $\gamma $(MeV) & $\delta $(MeV)
& $\varepsilon $($\frac{\text{c}^2}{\text{GeV}^2}$) \\ \hline
S & 200 & -356 & 303 & 1.17 & --- & --- \\ \hline
SM & 200 & -390 & 320 & 1.14 & 1.57 & 500 \\ \hline
H & 380 & -124 & 71 & 2.00 & --- & --- \\ \hline
HM & 380 & -130 & 59 & 2.09 & 1.57 & 500 \\ \hline
\end{tabular}
}
\end{center}

\section*{Figure captions}

\begin{description}
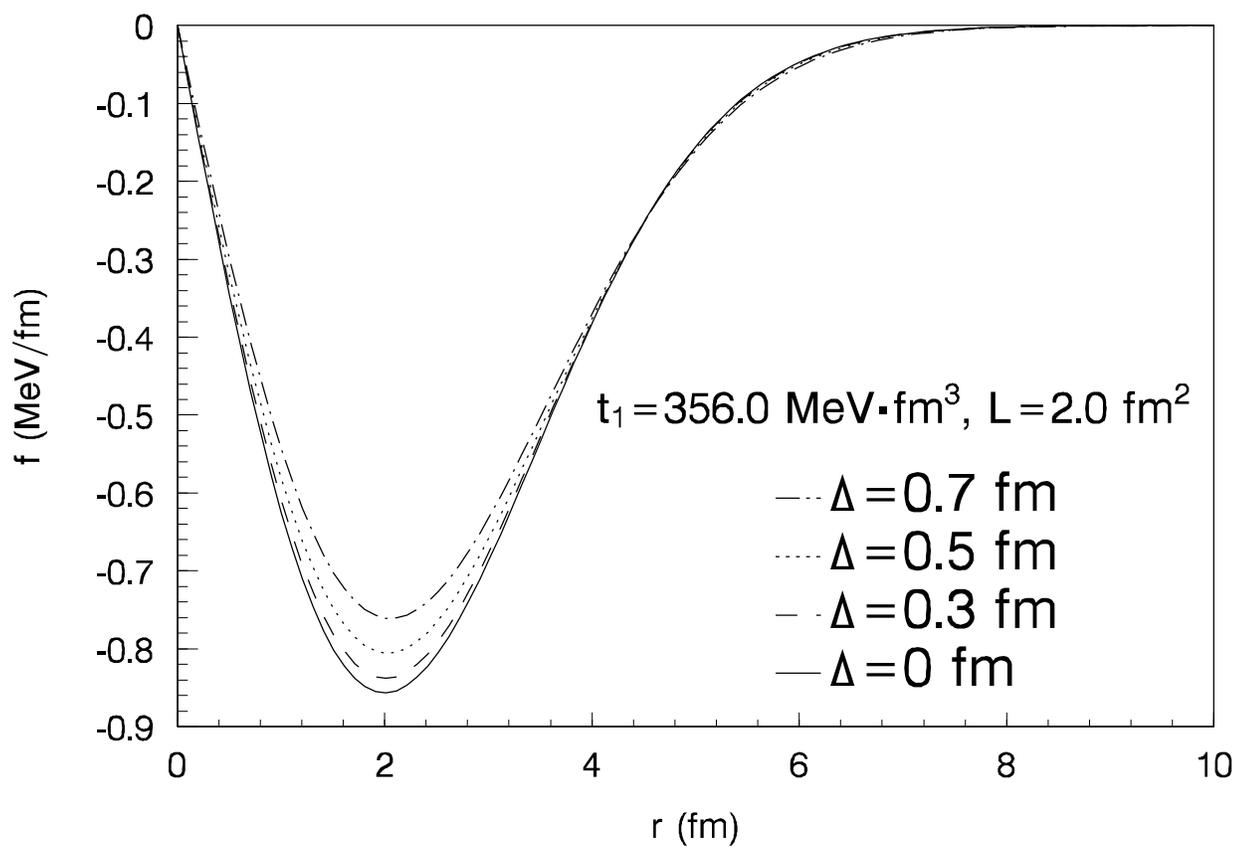
\item[FIG. 1]  The finite-range Gaussian 2-body nuclear force as a function
of the distance between the two nucleons for $\Delta =0.0$, $0.3$, $0.5$ and 
$0.7$ fm.

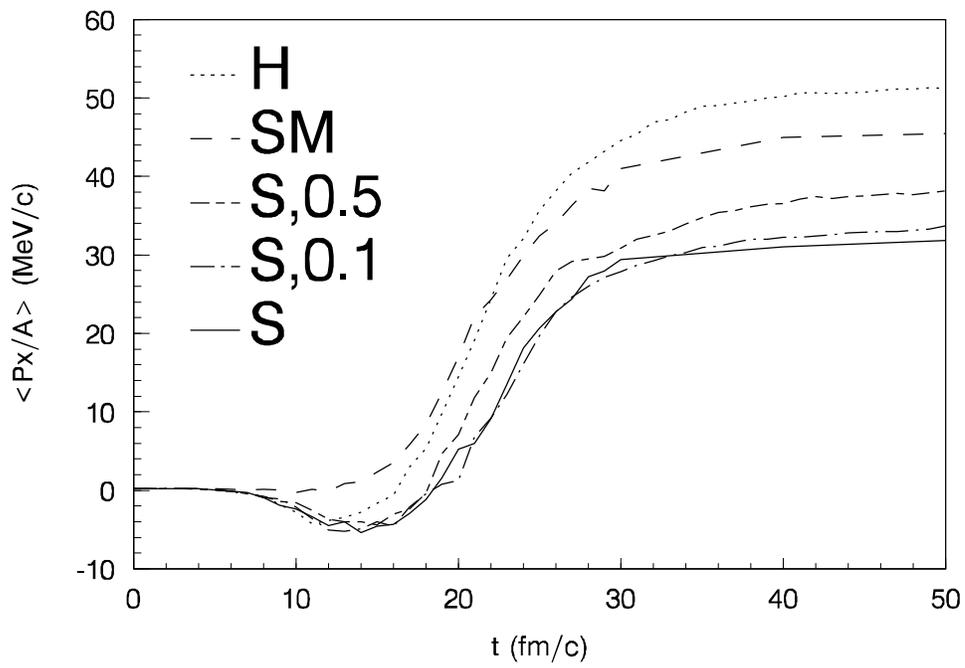
\item[FIG. 2]  Time evolution of transverse momentum per nucleon for five
different potential parameter sets, namely, hard EOS (H), soft EOS (S), soft
EOS with momentum dependent interaction (SM), soft EOS with $\Delta =0.1$ (S
0.1), and soft EOS with $\Delta =0.5$ (S 0.5) for the reaction $^{93}$Nb + $%
^{93}$Nb at energy of 400 MeV/nucleon and impact parameter $b$=3 fm.

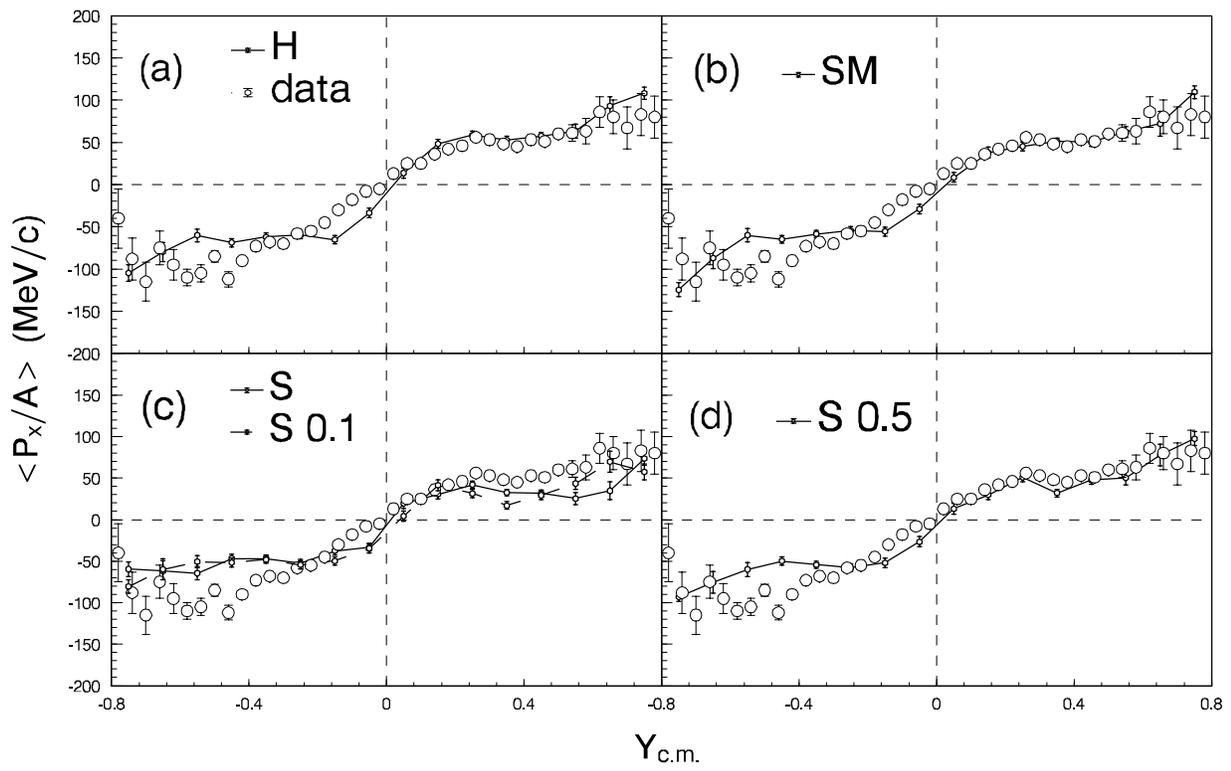
\item[FIG. 3]  Rapidity distribution of the transverse momentum for five
different potential parameter sets, namely, hard EOS (H) (a), soft EOS with
momentum dependent interaction (SM) (b), soft EOS (S) and soft EOS with $%
\Delta =0.1$ (S 0.1, the small open circles linked by dashed line) (c), and
soft EOS with $\Delta =0.5$ (S 0.5) (d). The larger open circles are the
experimental data and the small open circles are the calculated results.

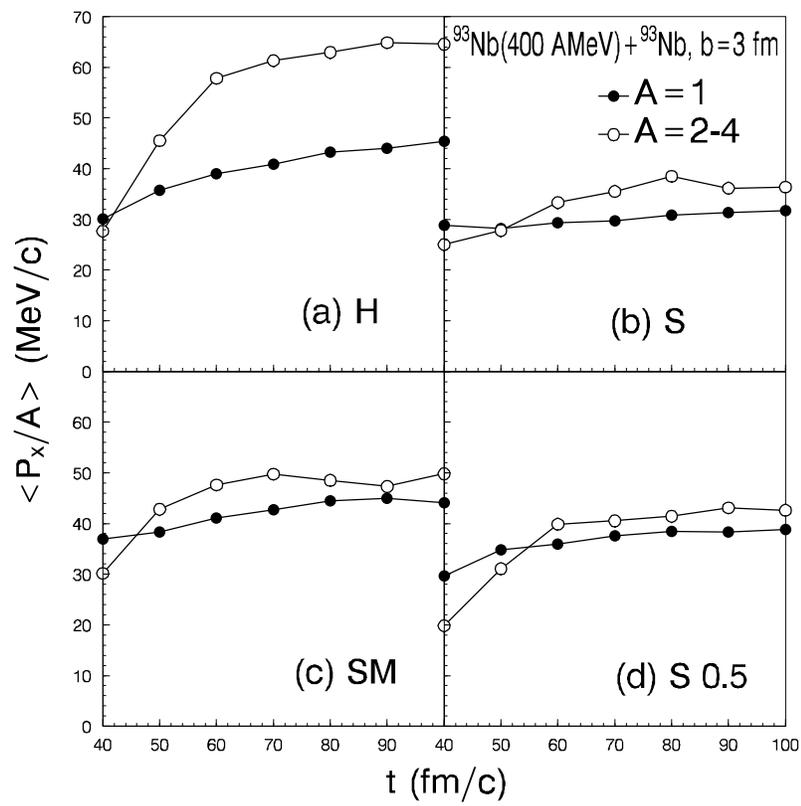
\item[FIG. 4]  Time evolution of transverse momentum per nucleon for free
nucleons (A=1, solid circles) and light fragments (A=2$\sim $4, open
circles) for four different cases, namely, hard EOS (H) (a), soft EOS (S)
(b), soft EOS with the momentum dependent interaction (SM) (c), and soft EOS
with $\Delta =0.5$ (S 0.5) (d). The lines included only to guide the eye.
\end{description}

\end{document}